\documentclass[twocolumn,showpacs,preprintnumbers,amsmath,amssymb,prb,aps]{revtex4}

\usepackage{graphicx}
\usepackage{dcolumn}
\usepackage{bm}
\usepackage{epstopdf}
\usepackage{amsmath}
\begin{document}

\title{Direct evidence of the existence of Mn$^{3+}$ ions in MnTiO$_3$}
\author{R. K. Maurya$^1$}
\author{Priyamedha Sharma$^1$}
\author{Ashutosh Patel$^2$}
\author{R. Bindu$^1$}
\altaffiliation{Corresponding author: bindu@iitmandi.ac.in}
\affiliation{$^1$School of Basic Sciences, Indian Institute of Technology Mandi, Kamand, Himachal Pradesh- 175005, India\linebreak
$^2$School of Engineering, Indian Institute of Technology Mandi, Kamand, Himachal Pradesh- 175005, India}

\date{\today}
\begin{abstract}
We investigate the room temperature electronic properties of MnTiO$_{3}$ synthesised by different preparation conditions. For this purpose, we prepared MnTiO$_{3}$ under two different cooling rates, one is naturally cooled while the other is quenched in $\emph{liq}$.nitrogen. The samples were studied using optical absorbance, photoemission spectroscopy and band structure calculations.We observe significant changes in the structural parameters as a result of quenching. Interestingly, in the parent compound, our combined core level, valence band and optical absorbance studies show the evidence of Mn existing in both 2+ and 3+ states. The fraction of Mn$^{3+}$ ions has been found to increase on quenching. The increase in the fraction of the Mn$^{3+}$ ions has been manifested (a) as slight enhancement in the intensity of the optical absorbance in the visible region.There occurs persistent photo-resistance when the incident light is terminated after shining; (b) in the behaviour of the features (close to fermi level) in the valence band spectra. Hence, the combined analysis of the core level, valence band and optical absorbance spectra suggest the charge carriers are hole like which further leads to the increase in the electrical conductivity of the quenched sample. The present results provide recipe to tune the optical absorption in the visible range for its applications in optical sensors, solar cell, etc.

\end{abstract}


\maketitle
Multiferroic materials find its application for various fields like high density data storage, solar cell, optical, humidity and magnetic sensor applications, etc\cite{nature,mrs,wang,acoustics,Tokura_apl}. These properties arise because of the intricate coupling between charge, spin, orbital and lattice degrees of freedom. MnTiO$_{3}$ is one such system which has shown to exhibit multiferroic properties along a particular crystallographic axis \cite{Mufti}.The systems under study belong to ilmenite family and stabilises in hexagonal (layered) structure with $\emph{R}$$\bar{3}$ space group down to low temperatures \cite{Mourya}. At room temperature, this material is paramagnetic.

The compounds under study also find its application in the field of solar cells \cite{solar} owing to its strong absorption of light in the visible region. In MnTiO$_{3}$, it has been predicted  \cite{seebeck} that conductivity mechanism is of p-type. Based on the high temperature (300 to 500 K) conductivity, Seebeck coefficient experiments, it was attributed that the mechanism could be due to the presence of higher oxidation state of Mn like Mn$^{3+}$. Apart from this, based on the value of the effective magnetic moment (4.55 $\mu$$_{B}$) obtained for Mn ion in MnTiO$_{3}$, the value is smaller than that of spin only value of Mn$^{2+}$. It was predicted that there could be incomplete ordering of Mn and Ti ions or the presence of Mn$^{3+}$ ions in this compound \cite{ND}.

\begin{figure}
\vspace{-10ex}
\includegraphics [scale=0.4, angle=0]{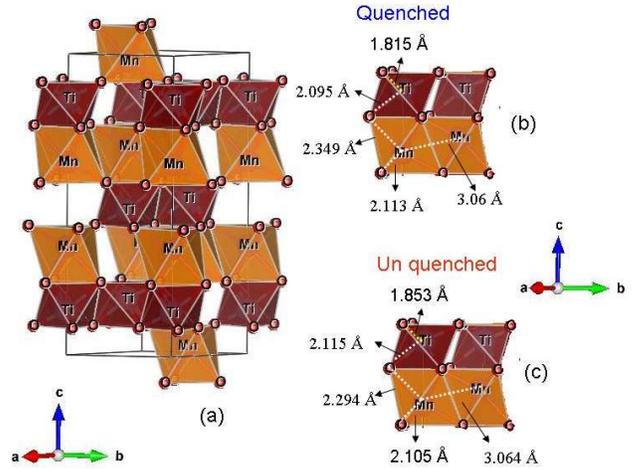}
\vspace{-25ex}
\caption{(a) The crystal structure of MnTiO$_{3}$. The local Mn-O, Ti-O and Mn-Mn (intra) bond distances for (b) quenched and (c)unquenched samples.}
\vspace{-2ex}
\end{figure}

Apart from this, if there is Mn$^{3+}$ existing in this compound, it is expected to be reflected in the core level spectra through the behaviour of the satellite peaks. Signature of such states is also expected to be manifested in the optical absorption spectrum of the MnTiO$_{3}$. The peak observed in the visible region has been attributed to intra-shell 3$\emph{d}$ transition of the transition metal ion and the one in the ultra violet (UV) region is linked with inter-band transition \cite{optical}. To understand its implications on the optical, transport and electronic structure, we have synthesised MnTiO$_{3}$ under different conditions and studied the above properties using x-ray diffraction, optical absorbance spectroscopy, photoemission spectroscopy and band structure calculations. Two compounds of MnTiO$_{3}$ were prepared by usual solid state route only with the difference that in one of the sample, the cooling was natural (MTO$\emph{uq}$) while in the other it was quenched (MTO$\emph{q}$) in $\emph{liq}$.nitrogen.

\begin{figure}
 \vspace{-1ex}
\includegraphics [scale=0.4, angle=0]{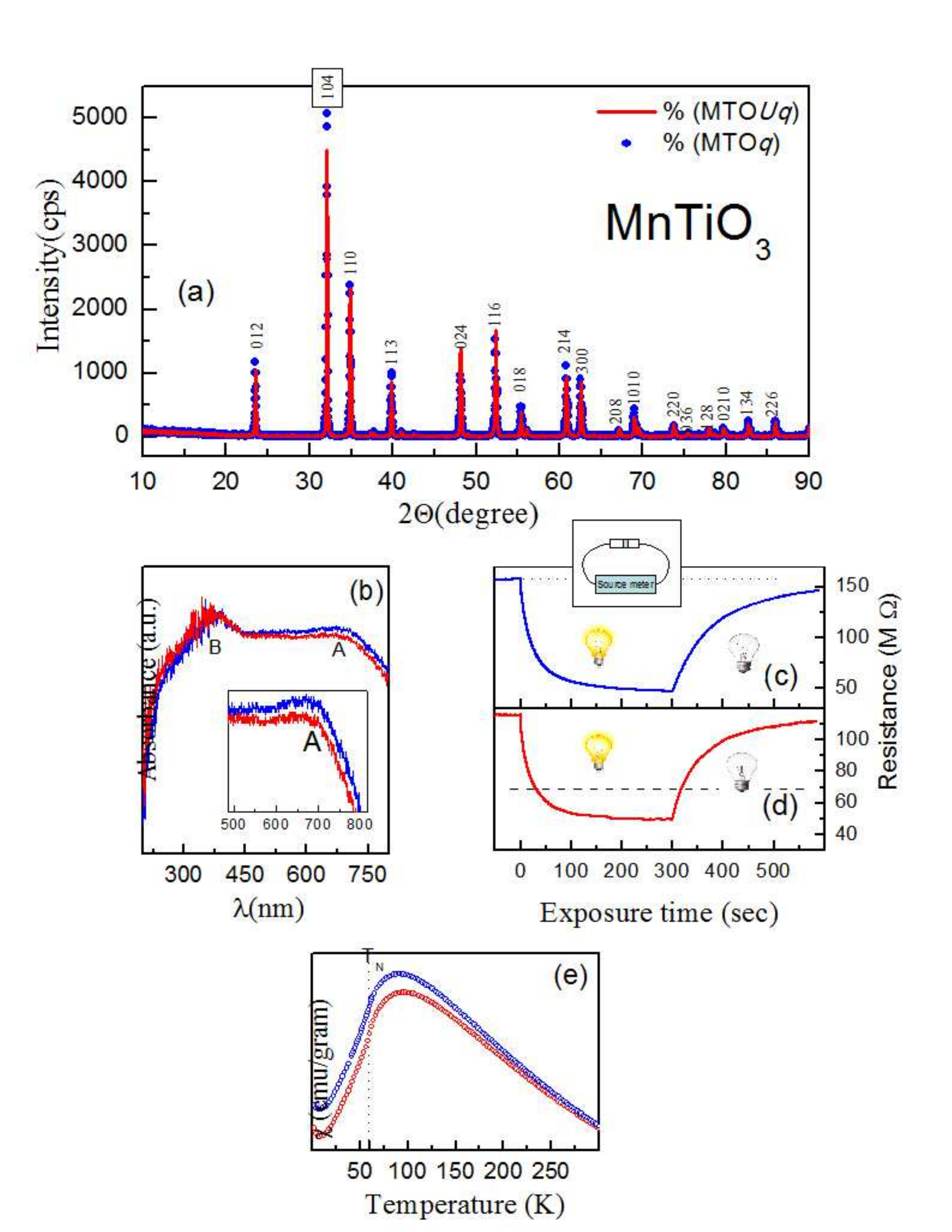}
 \vspace{-5ex}
\caption{Comparison of the xrd patterns of MTO$\emph{q}$ (solid circle) and MTO$\emph{uq}$ (solid line) samples. (b) The room temperature optical absorbance spectra of MTO$\emph{uq}$ and MTO$\emph{q}$. The inset shows the absorption spectra in the visible region. The resistance change after switching ON and OFF the white light (c)MTO$\emph{q}$ and (d) MTO$\emph{uq}$ samples. The inset shows the set up.(e) Comparison of the dc susceptibility of MTO$\emph{uq}$ (red open circle) and MTO$\emph{q}$ (blue open circle) carried out at applied field of 0.1 T.}
\vspace{-2ex}
\end{figure}

Our results show that on quenching the sample, there occurs, changes in the structural parameters, optical absorbance, core level and valence band spectra and the electrical resistivity.All the experiments were carried out at room temperature. We have observed the signature of Mn$^{3+}$ ions in addition to Mn$^{2+}$ ions even in the MTO$\emph{uq}$. The fraction of the Mn$^{3+}$ ions has been found to increase on quenching the sample. This further leads to the increase in the optical absorbance in visible region, persistent photo resistance when the incident light is terminated after illuminating it and decrease in the electrical resistivity.The features of the optical absorbance and valence band spectra were identified based on band structure calculations.

\begin{figure}
\vspace{-5ex}
\includegraphics [scale=0.45, angle=0]{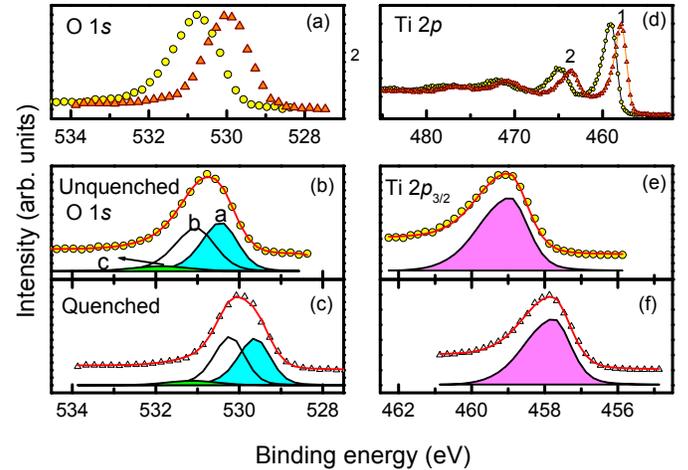}
\vspace{-40ex}
\caption{(a) Comparison of the O 1$\emph{s}$ spectra of MTO$\emph{uq}$ (open circles) and MTO$\emph{q}$ samples (open triangles);The fit (line)to the O 1$\emph{s}$ experimental spectra (symbol) of (b) MTO$\emph{uq}$ and (c) MTO$\emph{q}$; (d)Comparison of the Ti 2$\emph{p}$ spectra of MTO$\emph{uq}$ (open circles) and MTO$\emph{q}$ samples (open triangles); The fit (line) to the Ti 2$\emph{p}$$_{3/2}$ experimental spectra (symbol) of (e) MTO$\emph{uq}$ and (f) MTO$\emph{q}$;}
\vspace{-1ex}
\end{figure}

The samples were prepared by conventional solid state route. The starting materials; MnCO$_{3}$ and TiO$_{2}$ were ground using mortar and pestle and sintered at 1200 $^{\circ}$C for 24 hours in air. After this, we observed unreacted TiO$_{2}$ peaks in the x-ray diffraction pattern. Suitable amount of extra MnCO$_{3}$ was added and sintered until single phase compound was formed. After the sintering process, the sample was allowed to cool naturally. This constitutes MTO$\emph{uq}$ sample.To prepare the MTO$\emph{q}$ sample, the pellets of MTO$\emph{uq}$ were sintered at 1200 $^{\circ}$C for 1 hour and quenched in $\emph{liquid}$ nitrogen.

The samples were characterized using powder x-ray diffraction (xrd) technique using Smart lab 9 kW rotating anode x-ray diffractometer. Within the detection limit of XRD, no traces of secondary phases were found in the xrd patterns.Temperature dependent DC susceptibility measurements were carried out using MPMS set up in the temperature range (300K to 4 K) at an applied field of 0.1 T. The Neel temperature (T$_{N}$) for both the samples is around 60 K.

The absorbance spectra were collected using UV-Vis spectrophotometer (Shimadzu-2450) with a deuterium lamp in the wavelength range of 200 nm to 800 nm. The spectrophotometer was calibrated using Barium Sulphate powder. The photocurrent measurements were performed by depositing Aluminium electrode patterns through shadow mask by thermal evaporation deposition technique.  The thickness and the electrode separation was about 150 nm and 100 $\mu$m, respectively. Ni contacts were made on the electrodes. Three Watt white LED light was used as a source of irradiation.

The room temperature photoemission experiments were performed using monochromatic AlK$\alpha$ (1486.6 eV) source, with an energy resolution of 400meV and Scienta Analyser (R3000). The sample surface was cleaned $\emph{in situ}$ by scraping with a diamond file. The binding energy was calibrated by measuring the Fermi energy of Ag in electrical contact with the sample.

The base pressure during the measurement was 5 x 10$^{-10}$ mbar.

The spin polarised calculations for MnTiO$_{3}$ were carried out by using $\emph{state of the art}$ full potential linearized augmented plane wave (FP-LAPW) method within local density approximation using elk code\cite{elk}. The calculations were carried out in the Hexagonal phase using the room temperature lattice parameters\cite{Mourya}. The muffin-tin radii used were 2.4, 1.95 and 1.46 a.u. for Mn, Ti and O, respectively.The convergence was achieved by considering 512 $\emph{k}$ points within the first Brillouin zone. The error bar for the energy convergence was set to be smaller than 10$^{-4}$ Hartree/cell.

The crystal structure of MnTiO$_{3}$ is shown in Fig.1. In Fig.2(a), we show room temperature x-ray diffraction patterns of MTO$\emph{q}$ and MTO$\emph{uq}$.  As compared to the unquenched sample, the quenched one exhibits shift in the position of the xrd peaks and also variation in its relative intensities. This suggests changes in the lattice parameters and also in the atomic positions.These changes can be attributed to the freezing of the high temperature phase as a result of quenching. Both the xrd patterns were indexed using $\emph{R}$$\bar{3}$ space group.The structural parameters were obtained using Rietveld profile refinement software\cite{Rietveld,Rietveld1}. The goodness of fit thus obtained is $\sim$ 1.4. The lattice parameters thus obtained for MTO$\emph{uq}$ and MTO$\emph{q}$ are $a$= 5.1363 and 5.135 ${\AA}$; $c$=14.2791 and 14.2810 ${\AA}$, respectively. It is interesting to note that the volume of the MnO$_{6}$ octahedra has increased but the TiO$_{6}$ octahedra has decreased with quenching, Figs. 1(b and c). Keeping this in mind, one may expect significant changes in its electronic structure and hence its magnetism, multiferroic properties and optical properties etc. To observe the effect of such structural changes in the optical properties, optical absorbance experiments were carried out.

\begin{figure}
\vspace{-5ex}
\includegraphics [scale=0.42, angle=0]{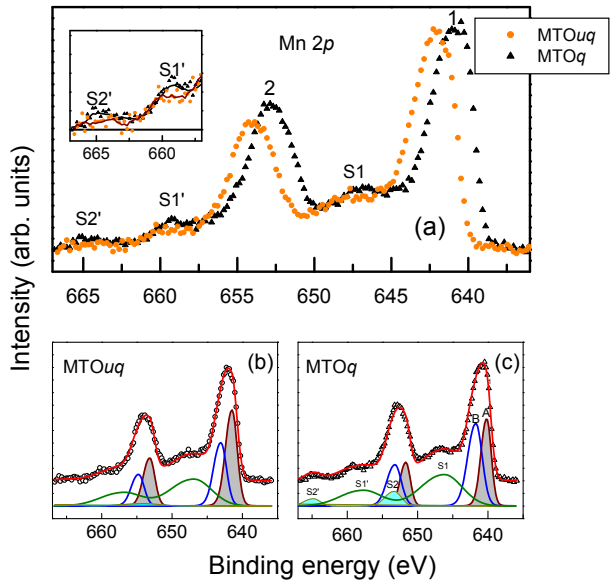}
\vspace{-25ex}
\caption{(a) comparison of the Mn2$\emph{p}$ of MTO$\emph{uq}$ (solid circle) and MTO$\emph{q}$ (solid triangle); The fit (line) to the experimental spectra of (b) MTO$\emph{uq}$ (open circle) and (c) MTO$\emph{q}$ (open triangles).}
\vspace{-4ex}
\end{figure}

Fig.2(b) shows the normalised optical absorbance spectra collected for both the samples. The optical absorbance was normalised by the area under the curve.  The absorbance spectra show significant contribution to the wide energy range covering UV and visible regions. Two broad peaks A and B centered around 695 nm (1.8 eV) and 365 nm (3.4 eV), respectively are observed. The quenched sample shows slight increase in the absorbance in the visible region while in the UV region it remains almost the same as the unquenched one.

Figs. 2(c and d) show the behaviour of the surface resistance as a function of exposure time when the samples were irradiated with white light. In the case of MTO$\emph{q}$, in the dark condition, the surface resistance was about 150 M$\Omega$ and reduced to 50 M$\Omega$ when the white light was switched on. In the case of MTO$\emph{uq}$, resistance reduced from 115 M$\Omega$ to 50 M$\Omega$. Such increased change in the resistance exhibited by MTO$\emph{q}$ is mainly because of increased optical absorbance as compared to MTO$\emph{uq}$ sample. When the white light illumination is withdrawn, the increase in the resistance consists of two components. In the beginning, it is fast and later on it is slow. It takes about 15 minutes to reach the dark resistance value.  This shows its ability to exhibit persistent photo-resistance. This means that there occurs some trap states that leads to the increase in the lifetime of the excited carriers. To understand the origin of such trap states, more experiments need to be performed.

To understand the origin of the behaviour of peak A, its important to understand (a) the valence state of the elements present in the compound; (b) and also the behaviour of the valence band.The above points can be captured based on the behaviour of the binding energy and the satellite features of the core levels and the valence band spectra. At room temperature, we have collected O 1$\emph{s}$, Mn 2$\emph{p}$, Ti 2$\emph{p}$ core level spectra and also valence band spectra.In comparison to MTO$\emph{uq}$ sample, the peak position of the C 1$\emph{s}$ of MTO$\emph{q}$ is found to be shifted by 0.2 eV towards lower binding energy .

In upper panels of Figs. 3 and 4,  we show the room temperature core level spectra of O 1$\emph{s}$, Ti 2$\emph{p}$ and Mn 2$\emph{p}$, respectively. The peak position of the Mn$^{3+}$ (peak B) in the Mn 2$\emph{p}$$_{3/2}$ spectra of the MTO$\emph{uq}$ sample is shifted towards higher binding energy as compared to the samples that do not show charging\cite{bindu,rao}. The binding energy positions of the O 1$\emph{s}$ and Ti 2$\emph{p}$$_{3/2}$ peaks are in line with titanates\cite{titanates}. This suggests charging effect in this sample. The effect of photoionisation cross section is different in different core levels. Hence the above mentioned core levels will have different effect of charging. In the case of MTO$\emph{q}$ sample, all the core levels are shifted towards lower binding energy due to reduced charging. This is because the room temperature resistivity of MTO$\emph{q}$ (0.11 M$\Omega$ cm) is drastically reduced as compared to MTO$\emph{uq}$ (89.48 M$\Omega$ cm) sample. In general, when there is inhomogeneous charging, the line shape of the core levels are expected to be affected. But in the current study, the line shapes appear to be the same. We now study in detail the information obtained from the core levels keeping the above constraints in mind.

All the spectra exhibit multiple structures. The O 1$\emph{s}$ spectra are fitted with 3 peaks, Figs. 3(b-c). Peaks $\emph{a}$ and $\emph{b}$ constitute the contribution from the Mn-O and Ti-O bonds, respectively due to different Madelung potential and the weak feature $\emph{c}$ arise due to the impurities from the surface and grain boundaries. In Fig. 3(d), we show the comparison of the Ti 2$\emph{p}$ core levels. The peaks 1 and 2 are the spin orbit (SO) split Ti 2$\emph{p}$$_{3/2}$ and Ti 2$\emph{p}$$_{1/2}$, respectively. The peaks above 470 eV are the charge transfer satellite peaks\cite{ct}. The Ti 2$\emph{p}$$_{3/2}$ is fitted with one peak, Figs. 3(e-f) that constitutes av.Ti-O bonds of the TiO$_{6}$ octahedra. The satellite features of the core level spectra are a good indicator for the valence state. The satellite peaks observed in Ti 2$\emph{p}$ core level is similar to that observed in SrTiO$_{3}$\cite{ct}. Hence Ti in both the samples is in 4+ state.

The comparison of the Mn 2$\emph{p}$ spectra of both the samples is shown in Fig.4(a). Peaks 1 and 2 constitute the spin orbit split ones namely 2$\emph{p}$$_{3/2}$ and 2$\emph{p}$$_{1/2}$. Interestingly, 3 satellite peaks are observed and labelled as s1, s1$'$ and s2$'$. The satellites s1 and s1$'$ are the ones corresponding to Mn 2$\emph{p}$$_{3/2}$ and 2$\emph{p}$$_{1/2}$, respectively of the Mn$^{2+}$ ions\cite{rao,Mn3+}. The satellite feature s2$'$ is an indication of the existence of Mn in 3+ state\cite{Mn3+}. The satellite corresponding to Mn 2$\emph{p}$$_{3/2}$ (Mn$^{3+}$) is hidden in Mn 2$\emph{p}$$_{1/2}$ peak. The one corresponding to Mn 2$\emph{p}$$_{1/2}$ is labelled as s2$'$. In the case of MTO$\emph{uq}$, s2$'$ is weak while in the case of MTO$\emph{q}$, it is predominant. This suggests that Mn exists in higher oxidation states other than Mn$^{2+}$, inset of Fig. 4(a) in both the compounds. In Figs.4 (b-c), we show the fitting of the Mn 2$\emph{p}$ spectra. The main peaks of the SO split ones are fitted with two peaks that constitute Mn$^{2+}$ and Mn$^{3+}$ states.The fitting of its corresponding satellites are also shown. Based on the fitting of the main peaks our results show the contribution of Mn$^{3+}$ ions in MTO$\emph{q}$ is more as compared to MTO$\emph{uq}$ sample.

\begin{figure}
\vspace{-5ex}
\includegraphics [scale=0.40, angle=0]{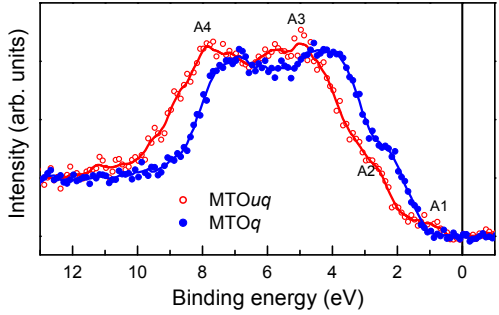}
\vspace{-45ex}
\caption{Comparison of the experimental room temperature valence band spectra of MTO$\emph{uq}$ and MTO$\emph{uq}$}
\vspace{-4ex}
\end{figure}

Fig. 5, shows the valence band (VB) spectra of both the samples. We observe the VB spectrum of MTO$\emph{q}$ is shifted towards lower binding energy with respect MTO$\emph{uq}$.  Here, zero binding energy marks the Fermi level ($\epsilon$$_{F}$). In the BE range 8 eV to $\epsilon$$_{F}$, the MTO$\emph{uq}$ sample shows 4 features while MTO$\emph{q}$ shows only 3 features. The feature A1 is suppressed in the latter case. To understand the behaviour of the features of the optical absorbance spectra and the VB spectra, band structure calculations were carried out, Fig.6. We now look into the results of band structure calculations to understand the optical and the valence band spectra.

The energy vs momentum ($\emph{k}$) curves for both the spin channels are given in Fig.6 (a).We do not observe any crossing of bands across the Fermi level ($\epsilon$$_{F}$) suggesting insulating ground state. The band gap obtained from the valence band maximum (VB$\emph{max}$) of spin up channel to the conduction band minimum (CB$min$) of the spin down channel is about 1 eV. The  VB$\emph{max}$ and the CB$min$ occur at two different symmetry points which is suggestive of indirect band gap semiconductor, Fig 6(a). In the VB region, close to $\epsilon$$_{F}$, Fig.6(a), we observe contribution of only spin up channel and in the CB region, the contribution is from the spin down channel only. In optical transitions, main contribution comes from direct band transitions. It is interesting to note that the band occurring around 1.3 eV is nearly dispersion less. This could be the possible reason for broad peaks observed in the optical absorbance spectra.

\begin{figure}
\vspace{-9ex}
\includegraphics [scale=0.43, angle=1800]{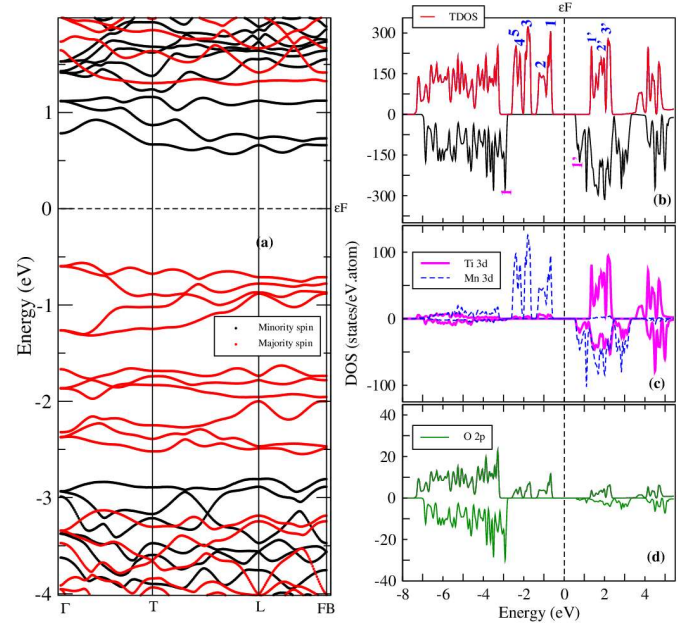}
\vspace{-20ex}
\caption{(a) Dispersion curves along different symmetry directions of the first Brillouin zone for both the spin channels. Fermi level is denoted by zero in the energy scale; (b)Total density of states (TDOS) corresponding to the up and down spin channels; Partial DOS corresponding to the up and down
spin channels of (c) Mn 3d, Ti 3d; (d) O 2p.}
\vspace{-2ex}
\end{figure}

Figs.6 (b-d) show the total density of states (TDOS) for up and down spin channels and the partial DOS corresponding to Mn 3$\emph{d}$, Ti 3$\emph{d}$ and O 2$\emph{p}$ states as a function of energy. In the VB region close to $\epsilon$$_{F}$, the contribution from Mn 3$\emph{d}$ states is dominant with less contribution from O 2$\emph{p}$. Around -0.6 eV, the hybridisation of Mn 3$\emph{d}$ with the O 2$\emph{p}$ states is more as compared to the region around -2.4 eV. There is also negligibly small contribution from Ti 3$\emph{d}$ states in these regions. On moving towards higher energy in the VB region, there is contribution from O 2$\emph{p}$, Mn 3$\emph{d}$ and Ti 3$\emph{d}$ states. In the CB region, there is significant contribution from Mn 3$\emph{d}$ and Ti 3$\emph{d}$ states with weak contribution from O 2$\emph{p}$ states.

From Fermi golden rule\cite{Chuang}, it is well known that absorption coefficient is directly related with the density of states. Keeping this in mind, transitions that contributes to the features in optical absorbance has been identified, Fig. 1(b). As per our calculations, the broad peak A is attributed to direct band transitions, Fig. 6(b) from 1 ($\sim$-0.62 eV) to 1' ($\sim$ 1.4 eV), 2'($\sim$ 1.94 eV) and 3' ($\sim$ 2.16 eV); 2 ($\sim$ -1.15 eV) to 1' ($\sim$ 1.4 eV). The broad peak B is attributed to the direct band transitions from 2 ($\sim$ -1.15 eV) to 2'($\sim$ 1.94 eV), 3' ($\sim$ 2.16 eV); 3 ($\sim$ -1.78 eV) to 1'($\sim$ 1.4 eV), 2' ($\sim$ 1.94 eV) and 3' ($\sim$ 2.16 eV) and also additional transitions from 6 ($\sim$ -2.9 eV) to 6'($\sim$ 0.82 eV). The increased intensity of peak B with respect to peak A could be due to these additional transitions. In the case of MnTiO$_{3}$, other reports \cite{optical}have attributed the peak A to intra-shell 3$\emph{d}$ transition of the Mn ion. While our calculations show that there is negligible probability of such transition. The increase in the intensity of peak A in MTO$\emph{q}$ as compared to MTO$\emph{uq}$ could be due to the increased probability of transition due to more number of available $\emph{d}$ states in the conduction band. This is expected to arise due to the increased fraction of Mn$^{3+}$ ions. This is manifested in the VB spectra in terms of decrease in the intensity of A1 feature.  The valence band spectra, Fig.5 exhibit a gap at the $\epsilon$$_{F}$ suggesting insulating nature of both the samples. Such behaviour suggests increase in the carrier concentration of holes in the MTO${q}$ sample. As a result the conductivity of the quenched sample is expected to be more than the unquenched one. This is in line with our resistivity results. We believe that our results will be useful in designing materials for optical sensors, and solar cell applications keeping in mind the cross link between the electronic structure, optical and transport properties.

In summary, we have studied room temperature structural, optical and electronic properties of quenched and unquenched polycrystalline MnTiO$_{3}$. Our results show that the quenched sample exhibits slight increase in the absorption in the visible region as compared to the unquenched one. In the UV region, the absorption of both the samples is identical. The features of the optical absorbance spectra have been identified based on the results of band structure calculations. The change in the surface resistance when illuminated with white light is more in the case of quenched sample as compared with the unquenched one. When the irradiated light is terminated, the samples show persistent photo-resistance. Based on the combined results of optical absorbance, core level, valence band studies, we find that Mn exists in 2+ and 3+ valence states even in the parent compound. The fraction of Mn$^{3+}$ ions increases on quenching the sample. Keeping in mind the above results and also the decrease in the room temperature electrical resistivity of the quenched sample suggests the charge carriers that participate in the conduction process are holes. Our results show direct link between the structural, electronic and optical properties and will help in better design of such materials for optical sensor and solar cell applications.

The authors thank Advanced Materials Research Centre, IIT Mandi for providing us the experimental facilities.

\end{document}